\documentclass[useAMS,usenatbib]{mn2e}

\usepackage{times}
\usepackage[latin1]{inputenc}
\usepackage{amssymb}
\usepackage{amsfonts}
\usepackage{amsmath}
\usepackage{mathrsfs}
\usepackage{txfonts}
\usepackage{graphicx}

\newcommand{\msol}{\ensuremath{M_\odot}}

\newcommand{\mdota}{\ensuremath{\dot{M}_{\rm a}}}
\newcommand{\LEdd}{\ensuremath{L_{\rm Edd}}}
\newcommand{\er}{\ensuremath{\epsilon_{\rm r}}}
\newcommand{\erad}{\ensuremath{\epsilon_{\rm r}}}
\newcommand{\ej}{\ensuremath{\epsilon_{\rm j}}}
\newcommand{\etot}{\ensuremath{\epsilon_{\rm a}}}
\newcommand{\sigmaT}{\sigma_{\scriptscriptstyle \rm T}}
\newcommand{\ltapprox}{\raisebox{-0.5ex}{$\,\stackrel{<}{\scriptstyle
\sim}\,$}}
\newcommand{\gtapprox}{\raisebox{-0.5ex}{$\,\stackrel{>}{\scriptstyle
\sim}\,$}}

\begin{document}


\title{Jet enhanced accretion growth of supermassive black holes}


\author[Jolley and Kuncic]{E. J. D. Jolley 
and Z. Kuncic \\
School of Physics, University of Sydney, NSW 2006, Australia\\
}


\date{}

\pagerange{\pageref{firstpage}--\pageref{lastpage}} \pubyear{2007}

\maketitle

\label{firstpage}

\begin{abstract}

We investigate the effect of a disc-driven 
jet on the accretion growth of cosmological supermassive black holes (SMBHs). The presence of a jet enhances the mass 
growth rate because for a given luminosity, the mass accretion rate, $\mdota$, is higher 
(or equivalently, the radiative efficiency $\er$ is lower for a fixed $\mdota$) 
than 
that predicted by standard accretion disc theory.
As jets carry away very little 
of the accreting matter, a larger proportion of the rest mass can reach the black hole during episodes of jet activity. 
We show quantitatively that the conditions required to grow a rapidly spinning black hole to a mass $\approx 10^9 \msol$ by redshift $z \approx 6$, 
whilst satisfying the observational constraint $\er \gtapprox 0.1$, are 
considerably less restrictive for jet-enhanced disc accretion than for standard disc accretion, which requires 
implausibly high super-Eddington accretion rates. 
Furthermore, jet-enhanced accretion growth offers a viable explanation for the observed correlation between
black hole mass and radio-loudness of quasars. 

\end{abstract}

\section{Introduction}

The discovery of over twenty quasars at 
redshift $z \gtapprox 5.7$ (e.g. \citealt{Fan00,Fan01a,Fan03,Fan04,Fan06a,Goto06}), with 
masses estimated to be $ \gtapprox 10^9 \msol$ 
\citep{Barth03, Vestergaard04, Jiang06a, Kurk07}, 
challenges our current understanding of the early universe at the epoch of galaxy formation. 
The highest redshift quasar, SDSS 1148+3251 \citep{Fan03}, with 
$z \approx 6.43$ (corresponding to a cosmic time $t=0.87$ Gyr assuming the standard 
$\Lambda$CDM concordance cosmology), is thought to harbour a black hole of mass 
$M_\bullet \approx \mbox{ a few } \times 10^9 \msol$ 
\citep{Fan03, Barth03, Willott03, Haiman04}. 
This poses a significant constraint on viable black hole growth mechanisms. 

In $\Lambda$CDM cosmology, 
dark matter halos merge heirarchically, with black holes merging and accreting 
in the gaseous centre (see e.g. \citealt{VolonteriRees05, Shapiro05, Hopkins06, Li07}). 
During mergers of host galaxies, the nuclear black holes can coalesce or be 
ejected due to gravitational wave recoil. Gravitational recoil does not significantly impede black hole growth 
if accretion is Eddington-limited \citep{Haiman04, YooMiralda04, Volonteri07}.
Coalescence should produce 
a linear relation between the halo mass and the black hole mass (see e.g. \citealt{Haenhelt98}). 
However, the observed relation between 
the black hole mass and the bulge velocity dispersion of galaxies \citep{FerrareseMerrit00} 
suggests instead that ongoing, sustained accretion, rather than coalescence as a result of mergers, 
is the primary process governing black hole growth. 

An important constraint on accretion growth models is that the mean 
radiative efficiency $\erad$ must be sufficiently low to 
maximize the amount of matter accreted onto the black hole and minimize the amount of rest mass energy lost to radiation. 
However, the observed ratio of the 
quasar/AGN luminosity density to the local SMBH mass density (at redshift $z \ltapprox 5$) 
requires $\er \gtapprox 0.1$ 
\citep{Soltan82,YuTremaine02,Elvis02, AllerRichstone02, Marconi04}. This is an 
empirical, model-independent lower limit. 
Since standard disc accretion onto a Schwarzschild black hole has a radiative 
efficiency $\erad \approx 0.06$ \citep{NovikovThorne}, it has been argued 
(e.g. \citet{Elvis02, YuTremaine02, Barger05}) that the 
Soltan relation implies that quasars harbour Kerr black holes. 
Indeed, 
models of cosmolgical black hole evolution \citep{Thorne74, Volonteri05} and merger-driven accretion 
\citep{diMatteo05}, constrained by 
the evolution of the luminosity function of quasars (see e.g. \citet{Hopkins07}), indicate that quasars 
undergo rapid spin-up. Furthermore, 
these models show that in order to grow cosmological black holes 
from an initial seed mass to that of a quasar by $z \gtapprox 3$, 
the amount of material accreted in a single 
accretion episode must be quite large. This acts to rapidly spin-up the 
black holes of massive high-redshift quasars \citep{VSL07}, 
contrary to the suggested low-spin scenario. We note, however, that 
models for low-spin black holes have also been 
proposed \citep{King05, KingPringle06}. 

The radiative efficiency of a maximally spinning accreting black hole 
is $\approx 40\%$, which is too high to grow a 
SMBH by $z \approx 6$, unless implausibly high, super-Eddington accretion rates are invoked. 
Observations indicate that the dimensionless mass accretion rate, $\dot{m} \equiv L / \LEdd$, 
where $\LEdd = 4 \pi G M_\bullet \mu m_{\rm p} c / \sigma_{\rm T}$ is the Eddington luminosity, 
is limited to $\dot{m} \ltapprox 2$ 
at redshifts approaching $z \approx 6$ \citep{Hopkins06}. Thus, it is difficult to reconcile 
standard disc accretion growth of 
cosmological, spinning SMBHs with the observational constraints $\er \gtapprox 0.1$ 
and $\dot{m} \simeq 1$. 

Mergers produce large-scale gravitational instabilities which can 
funnel gas down to the galactic core where it can be accreted \citep{MihosHernquist94a, MihosHernquist94b}. 
In order for efficient disc accretion to proceed, another process must 
facilitate the transport of angular momentum on smaller scales. Numerical 
simulations demonstrate that the effective viscosity produced by magnetohydrodynamical 
turbulence, generated by the magnetorotational instability, is insufficient to account 
for the very high mass accretion rates inferred in the most powerful accreting sources, 
such as quasars \citep{Hawley00, StonePringle01, Hawley01, HawleyBalbus02, KingPringle07}, 
unless a large-scale, systematic poloidal field is present in the accretion flow 
\citep{SteinackerHenning01, Campbell03, KigureShibata05, Salmeron07}. In this case, 
a magnetized jet forms and the overall mass accretion rate increases considerably 
as a result of enhanced angular momentum transport and negligible mass loss \citep{KuncicBicknell04, 2007a, 2007b}. 
This creates auspicious conditions for efficient mass growth of black holes. 

Approximately $60\%$ of all AGN display outflow phenomena \citep{GangulyBrotherton07}. 
Relativistic outflows in particular are a characteristic feature of accreting SMBHs. 
X-ray observations 
of galaxy clusters indicate that substantial amounts of mechanical energy may be deposited into the intracluster 
medium by powerful AGN jets (see e.g. \citealt{Birzan04, Allen06, Fabian06, Rafferty06, Taylor06}). 
If some fraction of the total accretion power is converted to jet kinetic power, then $\erad$ is lower 
than that predicted by standard disc accretion for a given $\mdota$. 
Hence, the black hole growth rate for jet-enhanced accretion is larger. 
Existing accretion growth models simply set the total accretion efficiency $\etot$ equal to 
the radiative efficiency $\er$ and thereby do not take into account 
the conversion of accretion power into 
non-radiative form (e.g. mechanical energy). 
Relativistic jets, in particular, can carry away a substantial amount of kinetic power but very little mass. 
Furthermore, the positive correlation between 
radio-loudness and black hole mass 
\citep{Laor00, Lacy01, McLureDunlop02, Oshlack02, WooUrry02, Dunlop03, 
Marziani03, Shields03, McLureJarvis04, MetcalfMagliocchetti06, Liu06}, 
suggests that jet-enhanced accretion 
growth results in more massive black holes. 

In this paper, we quantitatively determine the effect of jet-enhanced accretion 
on the cosmological growth of SMBHs. In 
Section \ref{sectionacc} we present results for a jet-enhanced 
accretion growth model. 
We discuss the implications of these results in \ref{sectiondisc} 
and give concluding remarks in Section \ref{sectionconc}. 

\section{Jet-Enhanced Accretion}\label{sectionacc}

The black hole mass growth rate is given by (see e.g. \citealt{Shapiro05}) 
\begin{equation}\label{dmdt}
\frac{dM_{\bullet}}{dt} = (1-\etot)\mdota
\end{equation}
where $\etot$ is the total accretion efficiency and $M_\bullet$ is the black hole mass. 
The efficiency of conversion of rest mass energy to radiation is 
\begin{equation}\label{erad}
\er = \frac{L} {\mdota c^2}
\end{equation}
where $L$ is the luminosity. 
Existing accretion growth models simply set $\er = \etot$ and hence do not take into account 
the conversion of accretion power into 
non-radiative forms, such as kinetic power in a relativistic jet. 
The total accretion efficiency should thus be more accurately expressed as 
\begin{equation}
\etot =  \er + \ej
\end{equation}
where $\ej$ is the jet efficiency (see \citealt{JolleyKuncic07}). 

Accretion power can be converted to jet power via 
a magnetic torque that acts over the 
disc surface and vertically transports angular momentum and energy 
from the disc \citep{KuncicBicknell04}. The rate at which work is done 
against the disk by the magnetic torque 
is \citep{JolleyKuncic07} 
\begin{equation}
P_{\rm j} = \frac{3}{2} c \int_{r_i}^\infty f_1(r) 
\left[ \int_{r_i}^r f_2(r) B_\phi B_z
\, \rm{d}r \right]\, \rm{d}r 
\end{equation}
where $f_1(r)$ and $f_2(r)$ are dimensionless functions of the disk radius, $r_i$ is the innermost stable orbit, 
and $B_\phi$ and $B_z$ are the azimuthal and vertical components of the magnetic field, respectively. 
Note that a large scale poloidal field is required to produce jets in numerical simulations 
(e.g. \citealt{SteinackerHenning01, Campbell03, KigureShibata05}).

A non-zero $\ej = P_{\rm j}/\mdota c^2$ can enhance accretion growth because jets transport angular momentum and thus 
give rise to a higher mass accretion rate \citep{KuncicBicknell04}. For a fixed $\etot$ (or a fixed $L$), this 
means that $\er$ is lower. 
The accretion efficiency $\etot$ for a relativistic disc \citep{NovikovThorne} 
evolves with the dimensionless spin $a = J/M_\bullet$ of the black hole, where $J$ is the 
specific angular momentum. 
The large amounts of material that must be accreted to form a $z \gtapprox 3$ quasar leads to 
very rapid 
spin-up of a black hole \citep{VSL07}. 
Thus, we take the spin $a$ and hence, the accretion efficiency \etot, to be approximately constant with time. 
If the average radiative efficiency and hence, average $\ej$, are also approximately constant, then 
the time evolution of black hole accretion growth is 
\begin{equation}
M_\bullet(t) = M_\bullet(t_0) \exp \left[ (t - t_0) \frac{(1-\etot)}{\er} \frac{4\pi G m_{\rm p}}{\sigmaT c} \dot{m} \right]
\end{equation}
where $M_\bullet(t_0)$ is the initial mass of the black hole. 
Cosmological simulations suggest 
seed black holes with $M_\bullet(t_0) \approx 600 \msol$ at $z \approx 25$ 
\citep{MadauRees01, OmukaiPalla03, Yoshida03}. 

\begin{figure*}
\centerline{\includegraphics[width=14truecm]{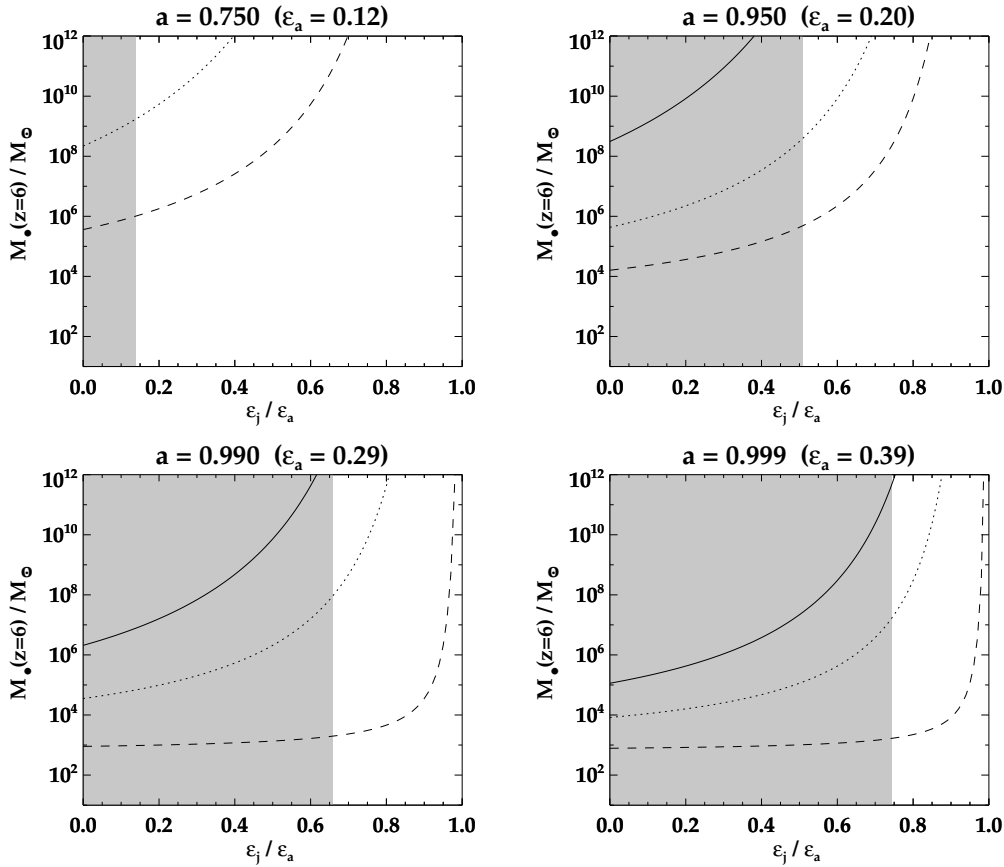}}
\caption{The effect of jet driven accretion on the 
final mass of a black hole at $z = 6$ growing from an initial mass 
$M_\bullet(t_0) = 600 \msol$ at redshift $z = 25$ 
for different values of the dimensionless spin $a$ and corresponding 
accretion efficiency $\etot$, as shown. 
The parameter $\ej/\etot$ is the fractional accretion power removed by a jet. 
The solid line is for an Eddington ratio $\dot{m} = 2$, the 
dotted line is for $\dot{m} = 1$ and the dashed line for 
$\dot{m} = 0.5$. The shaded areas indicate regions where the radiative efficiency is $\erad \ge 0.1$.}
\label{fig1}
\end{figure*}
%(a) has average spin $a=0.750$, corresponding to $\etot = 12\%$, 
%(b) has average spin $a=0.950$, corresponding to $\etot = 20\%$, 
%(c) has average spin $a=0.990$, corresponding to $\etot = 29\%$, and 
%(d) has average spin $a=0.999$, corresponding to $\etot = 39\%$. 
\begin{figure*}
\centerline{\includegraphics[height=20truecm]{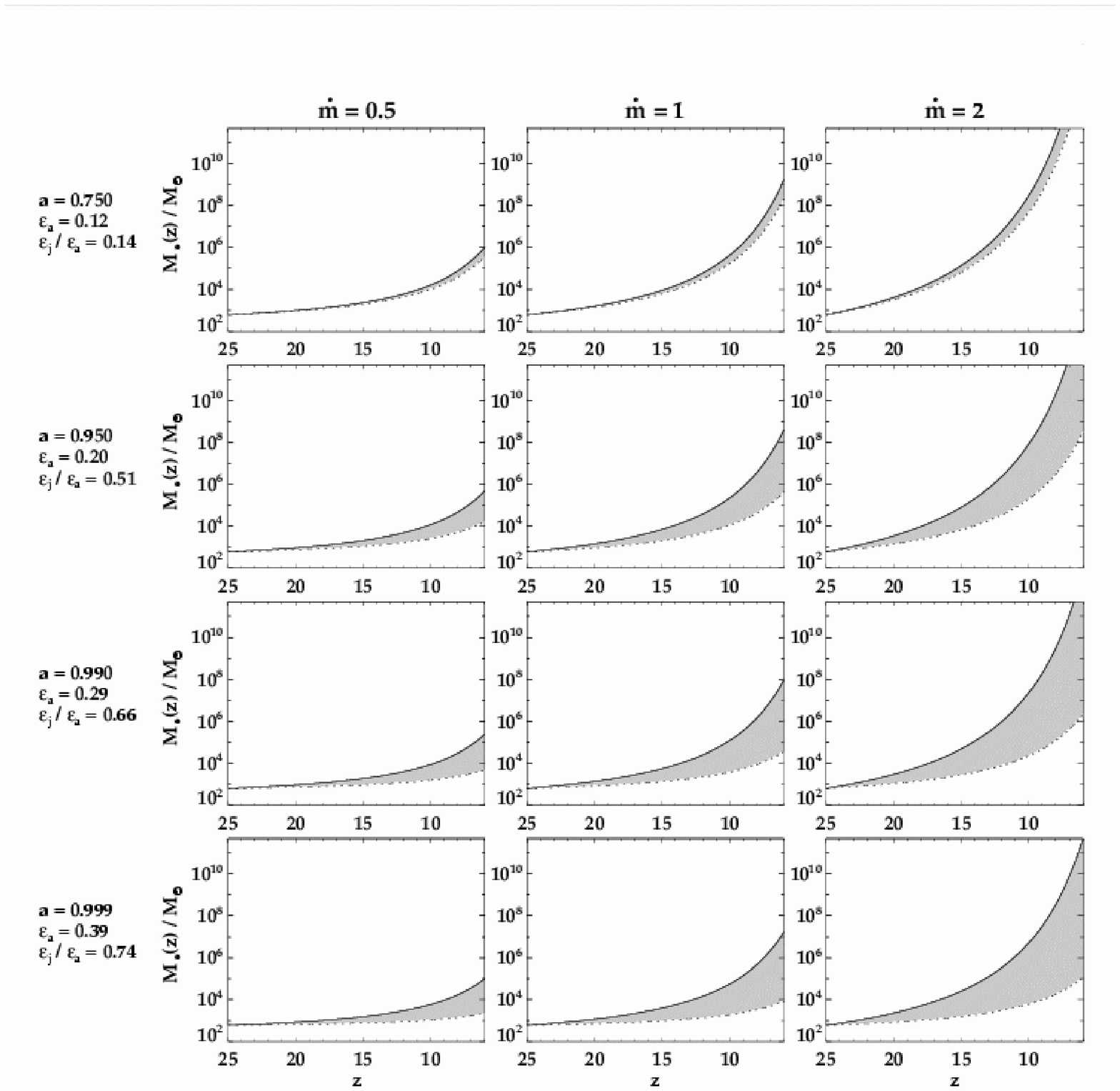}}
\caption{The black hole mass evolution as a function of redshift. The left column is for the 
case $\dot{m} = 0.5$, the middle column is for $\dot{m} = 1$, and the right column is for $\dot{m} = 2$. 
The top row is for a dimensionless spin $a=0.750$ (corresponding to an accretion 
efficiency $\etot = 0.12$), the second row is for 
$a=0.950$ ($\etot = 0.20$), 
the third row is for $a=0.990$ ($\etot = 0.29$) and the bottom row is 
for $a=0.999$ ($\etot = 0.39$). 
The solid line corresponds to the maximum possible value of $\ej/\etot$ such that 
$\erad = 0.1$, and the dotted line is the case without a jet i.e. $\ej/\etot = 0$ and $\erad = \etot$. 
The shaded areas indicate regions where the radiative efficiency is $\erad \ge 0.1$. }
\label{fig2}
\end{figure*}

Figure \ref{fig1} shows the dependence of the final mass of a black hole on the 
fraction of accretion power removed by a jet for different values of the spin $a$ and 
Eddington ratio $\dot{m}$. Standard disc accretion, which neglects vertical transport of angular momentum 
and energy by magnetized jets, 
corresponds to the case $\ej/\etot = 0$. 
The presence of a jet results in 
a higher final black hole mass than the standard disc accretion model. 
The shaded regions in Fig. \ref{fig1} indicate values of $\ej/\etot$ where $\erad \ge 0.1$, 
as implied by 
observations of the ratio of the AGN and quasar 
luminosity density to the SMBH mass density \citep{Soltan82}. 
Fig. \ref{fig1} shows that in order to grow a SMBH of mass $M_\bullet \approx 10^9 \msol$ 
by $z \approx 6$, a standard disc 
($\ej/\etot = 0$) requires a low spin, $a < 0.75$, or highly super-Eddington accretion, 
$\dot{m} \gg 2$. The presence of a jet significantly relaxes these constraints: 
Eddington-limited accretion can grow a rapidly spinning (with $a = 0.98$) 
black hole to $M_\bullet \approx 10^9 \msol$ 
by $z \approx 6$ and maintain a radiative efficiency 
$\er \approx 0.1$ if there is a jet that removes two-thirds 
of the accretion power. 

Figure \ref{fig2} shows the black hole accretion growth evolution as a function of $z$ for 
$25 \ge z \ge 6$ for different values of 
$\dot{m}$ and $a$. The solid curves are for jet-enhanced 
accretion growth with the largest possible fractional jet power $\ej/\etot$, 
corresponding to $\er = 0.1$. The dotted curves correspond to 
accretion growth via a standard disc (i.e. $\ej = 0$). The shaded region indicates where $\erad \ge 0.1$. 
Jet-enhanced accretion growth 
is evident in all cases, with the most enhancement occuring at high $a$ and $\dot{m} \gtapprox 1$. 
Note that for plausible physical parameters, black 
holes cannot grow to masses $\approx 10^9 \msol$ by $z \approx 6$ for sub-Eddington accretion rates 
even in the case of jet-enhanced accretion (Fig. \ref{fig2}; left column). 

For Eddington-limited accretion (Fig. \ref{fig2}; middle column), a black hole of $\approx 10^9 \msol$ 
at $z \approx 6$ can be formed for jet-enhanced accretion (solid line) provided $a \ltapprox 0.95$. 
By comparison, standard disc accretion at the Eddington rate (dashed line) requires 
an even lower spin $a \ltapprox 0.75$, 
which is difficult to reconcile with the rapid spin evolution of high-redshift
quasars \citep{VSL07}. 
For super-Eddington accretion with $\dot{m} =2$ (Fig. \ref{fig2}; right column), 
standard disc accretion still requires the spin to be below $0.95$ to grow a $10^9 \msol$ black hole by $z \approx 6$, 
whereas jet-enhanced accretion achieves this for all spins up to $a=0.999$. 
Clearly, the presence of a jet can substantially relax the requirements for growing cosmological SMBHs. 
The rapid evolution of black holes to a near-maximal spin \citep{Thorne74, Volonteri05} poses a serious problem 
to standard disc accretion, which cannot grow high-spin SMBHs rapidly enough without invoking implausibly high 
super-Eddington accretion rates. For example, $\dot{m} = 4$ is 
required by the standard model to grow a $M_\bullet \approx 10^9 \msol$ 
black hole with $a = 0.999$ by $z=6$. Jet-enhanced accretion requires just $\dot{m} = 1.3$, with 
two thirds of the accretion power in this case used to power a jet, thereby 
reducing the radiative efficiency to $\er = 0.1$. 

\section{Discussion}\label{sectiondisc}

The presence of a jet can greatly enhance the black hole mass growth rate because 
for a given mass accretion rate, the radiative efficiency is lower than that of 
a standard disc. Or equivalently, the accretion rate is higher for a given luminosity. 
This arises because jets enhance angular momentum transport; the additional accretion power, which depends on the accretion rate, drives the jet.
An important constraint for jet-enhanced accretion growth of black holes is the average jet efficiency 
over the lifetime of the rapid growth phase ($\Delta t \approx 0.7$ Gyr). 
Recent studies (see e.g. \citealt{Binney07, Kording07}) suggest relativistic jets can remove well in excess 
of half the total accretion power (i.e. $\ej/\etot \gtapprox 50\%$), implying they are an "all or nothing" phenomenon. 
When combined with the estimated $\approx 18\%$ duty cycle for quasar activity 
for $M_\bullet \gtapprox 10^9 \msol$ 
and $z \ltapprox 2$ (see \citealt{Wang06} and references therein), this suggests a lower limit to the 
fractional jet power of 
$\langle \ej \rangle / \etot \gtapprox 10\%$. For a black hole with $a = 0.99$ ($\etot \approx 0.29$), this gives 
$\langle \ej \rangle \gtapprox 0.03$, consistent with the value independently derived by \citet{Heinz07}.
This implies that, on average, accreting black holes liberate most of their energy in the form 
of radiation rather than jet kinetic energy. However, black holes grow  faster during jet-active 
phases because mass accretion is enhanced and we argue that such rapid growth episodes are necessary to 
explain the most massive cosmological black holes.

If the quasar duty cycle extrapolates to high $z$, then we predict jet-enhanced 
accretion to grow a $\approx 10^9 \msol$ 
black hole by $z \approx 6$ for $\dot{m} \ltapprox 3$. 
For comparison, standard disc accretion would require an accretion rate that is at least $30\%$ higher. 
Furthermore, this may be an overly conservative estimate if jets were more prevalent at high $z$ and hence, 
the quasar activity duty cycle was higher. Existing data suggest that jets were indeed more common 
at $z \approx 2$ than today, but the radio luminosity function at $z \gtapprox 2$ is still not well constrained 
(see e.g. \citealt{Willott01, LiuZhang07}). 

For a $\approx 10^9 \msol$ accreting black hole with $a = 0.99$ ($\etot \approx 0.29$) and $\langle \ej \rangle = 0.03$ the average accretion rate needed to produce a luminosity $L = 3 L_{\rm Edd} \approx 4 \times 10^{47} {\rm erg \, s}^{-1}$ with an average radiative efficiency $\langle \er \rangle =0.26$  is $\approx 25 M_\odot \, {\rm yr}^{-1}$.  The average jet power is 
$\langle P_{\rm j} \rangle \approx 4 \times 10^{46}\, \rm{ erg \,s}^{-1}$, 
which agrees well with jet kinetic powers 
inferred observationally (e.g. \citealt{MerloniHeinz07}; see also \citealt{CelottiGhisellini07} for a  
$z \approx 5.7$ jet). 
As jets have radiative efficiencies typically $\ltapprox 1\%$ (see e.g. \citealt{MerloniHeinz07}), virtually all 
the kinetic energy is deposited into the ISM or ICM. Over the rapid growth phase ($\Delta t \approx 0.7$ Gyr), 
the average total energy deposited is $\langle P_{\rm j} \Delta t \rangle \approx 10^{63}$ erg. 
This is consistent with the strong lower limit of $10^{60}$ erg inferred from X-ray cavities 
in rich cluster cores \citep{Birzan04, Binney07}. 

Finally, we can check whether spinning accreting black holes in AGN remain consistent 
with observations of the X-ray background (XRB) 
by matching the expected mass density of relic black holes 
deduced from the XRB (see \citealt{Marconi04} and references therein), 
\begin{equation}\label{rho}
\rho_{\rm XRB} = (4.7 - 10.6) \times \left(\frac{1-\langle \etot \rangle}{9\langle \er \rangle}\right)\times 10^5 \, \msol \, {\rm Mpc}^{-3}
\end{equation}
to the local black hole mass density $\rho_{\rm BH} = (4.4 - 5.9) \times 10^5 \, \msol \, {\rm Mpc}^{-3}$ 
\citep{GrahamDriver07}. \citet{Marconi04} show that $\rho_{\rm XRB}$ is compatible with $\rho_{\rm BH}$ without requiring radiative efficiencies considerably higher than $\sim 0.1$. Although they used an earlier estimate of $\rho_{\rm BH}$ with a slightly broader range, their result still holds for the more restrictive range deduced by \citet{GrahamDriver07}. However, as pointed out by \citet{Marconi04}, if some of the accretion energy emerges in non-radiative (i.e. mechanical) form (i.e. $\etot \neq \er$), then it is possible to place some constraints on the mean accretion efficiency $\langle \etot \rangle$ and thereby the mean spin of accreting black holes in the AGN population. We do not expect the mean black hole spin across the whole AGN population to be necessarily as high as that of the high-$z$ population. Indeed, $\rho_{\rm XRB}$ and $\rho_{\rm BH}$ can be compatible with each other and with the Soltan relation, $\langle \er \rangle \gtapprox 0.1$, if the mean accretion efficiency in the AGN population is $\langle \etot \rangle \ltapprox 0.2$, corresponding to black hole spins $a \ltapprox 0.95$. This also implies $\langle \ej \rangle \ltapprox  0.1$ and since this is comparable to the mean jet efficiency we deduced at high-$z$, it suggests a fundamental jet production mechanism that remains remarkably the same across the entire accreting SMBH population.

\section{conclusions}\label{sectionconc}

High redshift accreting black holes are  rapidly spinning and produce an enormous amount of power.
The ubiquity of jets and outflow phenomena amongst accretion-powered sources provides clear evidence that not all of 
the accretion power is radiated away, as predicted by standard accretion disc theory. 
That is, for a given luminosity, the mass accretion rate of sources with 
jets is higher than that of sources without jets. 
This is because jets enhance angular momentum transport while minimizing mass loss. 
We have shown that the conditions required to grow a rapidly 
spinning black hole to a final mass $\approx 10^9 \msol$ 
by $z \approx 6$ are considerably more easily met for 
jet-enhanced disc accretion than for standard disc accretion. 
Our results indicate that while accreting black holes, on average, liberate most of their energy in radiation, the most massive cosmological black holes may undergo rapid growth episodes associated with enhanced rates of mass accretion driven by intermittent jet activity.
This is consistent with the observed correlation between radio-loudness and 
black hole mass in AGN. 
It also implies that jets may have provided an 
important means of angular momentum and energy transport in the high-redshift 
universe. 

\section{acknowledgments}

E. J. D. J. acknowledges support from a University of Sydney Postgraduate Award.
The authors wish to thank Katherine Blundell, Roberto Soria and Geoff Bicknell for 
valuable discussions and an anonymous referee for useful comments that helped to improve the paper.

\label{lastpage}
\end{document}